# Scaling of the energy spectra of wall-bounded turbulence

J. C. del Alamo[1], J. Jiménez[1,2], P. Zandonade[3], R. D. Moser[3]

[1]School of Aeronautics, 28040 Madrid, Spain. [2]Center for Turbulence Research, Stanford, CA 94305. [3]TAM, University of Illinois, Urbana, IL 61801, USA.

Contact address: juanc@torroja.dmt.upm.es

## 1 Introduction

By using new results from direct simulations of turbulent channels at moderate friction Reynolds numbers ($Re_\tau \leq 1900$) and in very large numerical boxes [2], we examine the corrections to the similarity assumptions in the overlap and outer regions of wall-bounded turbulence.

The simulations are summarized in table 1. The spatial discretization uses Fourier expansions in the streamwise (x) and spanwise (z) directions, and Chebychev polynomials the wall-normal direction (y). The streamwise velocity component is u, its Fourier transform is denoted $\hat{u}$, and h is the channel half-height.

| Case  | $Re_\tau$ | $L_x/h$ | $L_z/h$ | $N_x$ | $N_z$ | $N_y$ |
|-------|-----------|---------|---------|-------|-------|-------|
| L550  | 547       | $8\pi$  | $4\pi$  | 1536  | 1536  | 257   |
| L950  | 934       | $8\pi$  | $3\pi$  | 3072  | 2403  | 385   |
| S1900 | 1901      | $\pi$   | $\pi/2$ | 768   | 768   | 769   |

Table 1: Parameters of the simulations. $N_x$ and $N_z$ are the number of collocation points in x and z, and $N_y$ is the number of Chebychev polynomials.

## 2 Results and discusson

The first deviation from simple self-similarity occurs in the width $\lambda_z$ of a range of u-structures, which scales as the square root of their length $\lambda_x = 2\pi/k_x$. This behaviour can be observed in the premultiplied two-dimensional spectrum of the streamwise velocity, $\Phi_{uu} = k_x k_z \langle \hat{u}\hat{u}^*\rangle$, which has been represented in figure 1(a), and suggests that the long modes of u act like a passive quantity dispersed by the smaller-scale background active turbulence. The range $\lambda_z \approx \lambda_x$ of figure 1(a) represents the linear dispersion of fluid elements for separations shorter than the integral scale of the active eddies. This scale is of the order of $\lambda_x = 10y$, and marks both the transition between the linear and square-root behaviours of $\Phi_{uu}$ (fig. 1a) and the long-wave cut-offs of $\Phi_{vv}$ and $\Phi_{ww}$ (ref. [2]), which agrees with the idea that blocking by the wall limits the size of the active wall-normal motions. Hence, most of $\langle uv \rangle$ is generated by the eddies in the range $\lambda_x < 10y$, suggesting that the intensity of the velocity fluctuations in that region should scale well with $u_\tau$, which is confirmed by the collapse of the data in figure 1(a).

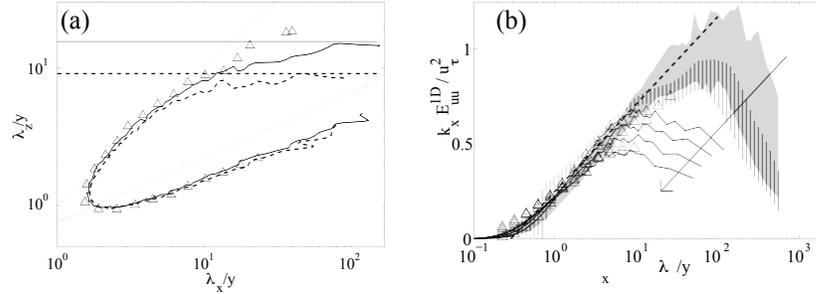

Figure 1: (a) Two-dimensional spectral density $\Phi^+_{uu}$ as a function of $\lambda_x/y$ and $\lambda_z/y$, $y^+ = 150$. The contours are $\Phi^+_{uu} = 0.2$. ——, L550; ——, L950; 6, S1900. The straight chaindotted line is $\lambda_x = \lambda_z$ and the dotted one is $\lambda_x^2 = \lambda_z y$. The dashed and solid straight lines are $\lambda_z = 2h$ for L550 and L950 respectively.
(b) Premultiplied one-dimensional spectra $k_x E^{1D+}_{uu}$ as functions of $\lambda_x/y$. The wall distance, $y^+ > 200$, increases in the sense of the arrow. Only $y/h < 0.1$ has been considered in the experiments to avoid contamination by the points in the outer layer, while up to $y/h = 0.5$ has been represented for the DNS's. The dotted straight line is $0.2 \log(4\lambda_x/y)$. The shaded area covers the maximum scatter of experimental boundary layers from ref. [6], $Re_\tau = 1300 - 7100$. The hatched regions are pipes from ref. [7], $Re_\tau = 2325 - 4900$.

A consequence of these scalings is a logarithmic correction to the $k_x^{-1}$ one-dimensional u-spectrum in the range $\lambda_x < 10y$, in which $\Phi_{uu}$ is limited by the square-root lower bound $\lambda_{z1} \approx \alpha^{-1}(\lambda_x y)^{1/2}$, and the linear upper bound $\lambda_{z2} \approx \lambda_x$ in figure 1(a). Integrating $\Phi_{uu}$ over $\lambda_z$ we obtain,

$$k_x E^{1D} = \int_{\lambda_{z1}}^{\lambda_{z2}} \Phi_{uu} \frac{d\lambda_z}{\lambda_z} \approx \beta u_\tau^2 \log(\alpha^2 \lambda_x/y), \qquad (1)$$

Figure 1(b) shows that (1) describes well the numerical and experimental spectra in the expected range. The two numerical coefficients $\alpha \approx 2$ and $\beta \approx 0.2$ in (1) have been chosen to fit the data in figure 1(b), but their values are also consistent with the intensity of $\Phi_{uu}$ and the position of its lower limit in figure 1(a).

A second non-classical scaling is the characteristic velocity of the largest u-structures, which are correlated across the entire flow [3, 1]. There is no reason why these global modes, which carry little Reynolds stress, should scale with $u_\tau$ and, because their large correlation heights suggest that they originate from the stirring of the whole velocity profile, from the wall to the top of the flow, a more plausible scale is the maximum mean velocity $U_c$. This is confirmed in figure 2(a), which displays the average $u'^2$ in the spectral band $6 < \lambda_x/h < 24$ as a function of Reynolds number. When it is scaled with $u_\tau^2$ it increases with $Re_\tau$ by approximately a factor of 2, but it remains roughly constant when scaled with $U_c^2$, except for the lowest Reynolds numbers.

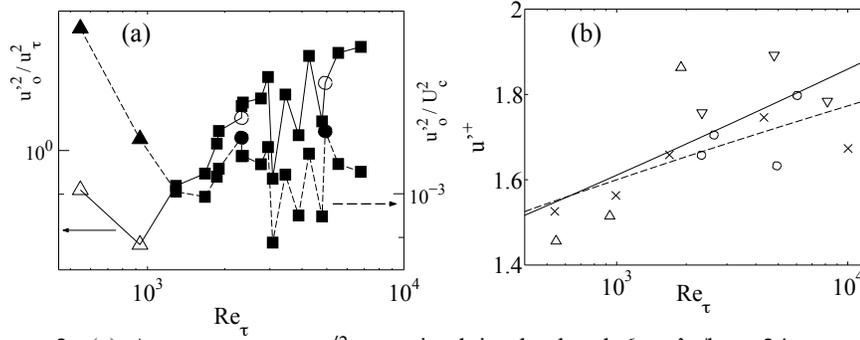

Figure 2: (a) Average energy $u'^2_0$ contained in the band $6 < \lambda_x/h < 24$ as a function of $Re_\tau$, $y/h = 0.075$. Open symbols, scaled with $u_\tau$ (left scale); solid symbols, scaled with $U_c$ (right scale). (b) Streamwise velocity fluctuations $u'^+$ as a function of $Re_\tau$, $y/h = 0.4$. ——, scaling proposed in eqs. 3 and 4; ——, mixed scaling proposed in ref. [5]. Channels: $\triangle$, present DNS; $\triangledown$, ref. [4]. Pipes: $\circ$, ref. [7]. Boundary layers: $\times$, ref. [5].

The maximum width of $\Phi_{uu}$ for the long global modes is limited to $\lambda_z = \lambda_{zc} \approx 2h$, as it can be observed for cases L550 and L950 in figure 1(a). This is the only limit in that figure which does not scale with $y$, and its effect is to "cut" $\Phi_{uu}$ progressively at lower values of $\lambda/y$ with increasing wall-distance. Computing the one-dimensional spectrum in the for the global modes, we obtain

$$k_x E^{1D}_{uu} \approx \int_{\lambda_{z1}}^{\lambda_{zc}} \Phi_{uu} \frac{d\lambda_z}{\lambda_z} \approx U_c^2 f(y/h) \log\left(\frac{\alpha^2 \lambda_{zc}^2}{\lambda_x y}\right), \qquad (2)$$

where the amplitude $f(y/h)$ depends on $y/h$ to reflect the the vertical structure of the global eddies.

As a result of the "cutting" effect of the limit $\lambda_z = \lambda_{zc}$, the intermediate region of the spectrum which separates the ranges of application of (1) and (2) disappears above $y > \lambda_{zc}/10 \approx 0.2 h$. This part of the flow is particularly easy to analyze, and the total energy of the fluctuations can be obtained by integrating (1) for $\lambda_x < \lambda_{zc}$ and (2) for $\lambda_x > \lambda_{zc}$. This results in

$$u'^2 \approx u_0^2 \log^2(\alpha \lambda_{zc}/y), \qquad (3)$$

where $u_0$ is a mixed velocity scale

$$u_0^2 = u_\tau^2 [\beta + f(y/h) U_c^{+2}]/2, \qquad (4)$$

which depends on the wall-distance, and on the Reynolds number through $U_c^+$. Figure 2(b) shows that this scaling predicts reasonably well the Reynolds number variation of the streamwise velocity fluctuations in the outer layer of boundary layers, pipes and channels. Because of the slow growth of $U_c$ with $Re_\tau$, the velocity scale 4 roughly coincides, in the range of Reynolds numbers considered in figure 2(b), with the one proposed by De Graaf and Eaton [5], $u'^{+2} \sim U_c^+$, but it tends to $u'^{+2} = f U_c^{+2}/2$ at much higher Reynolds numbers. Only when $U_c^+ \gtrsim 50$, roughly $Re_\tau \gtrsim 10^8$, the former scaling laws differ substantially.

## 3  Conclusions

The spectra of the velocity fluctuations in turbulent channels have been analyzed using direct numerical simulations with Reynolds numbers up to $Re_\tau = 1900$. Their scaling is anomalous in several respects, including a square-root behaviour of their width with respect to their length, and a velocity scaling of the largest modes with the centre-line velocity $U_c$. This implies a logarithmic correction to the $k^{-1}$ energy spectrum, and leads to a scaling of $u'$ away from the wall which agrees with the mixed scaling of by De Graaf and Eaton [5] at intermediate Reynolds numbers, but which tends to a pure scaling with $U_c$ at very large ones.

This work has been funded by CICYT, CEPBA/IBM (Spain), ONR, NSF, AFOSR and DOE ASCI (US).